\documentclass[1p]{elsarticle}

\usepackage{lineno}
\modulolinenumbers[5]
\usepackage{amssymb}
\usepackage[colorlinks=true,linkcolor=blue,urlcolor=blue,citecolor=blue]{hyperref}
\biboptions{numbers,sort&compress}

\begin{document}

\begin{frontmatter}

\title{Frustrated spin-1/2 ladder with ferro- and antiferromagnetic legs}

\author{Debasmita Maiti}

\author{Dayasindhu Dey}

\author{Manoranjan Kumar\corref{mycorrespondingauthor}}
\cortext[mycorrespondingauthor]{Corresponding author}
\ead{manoranjan.kumar@bose.res.in}
\address{S. N. Bose National Centre for Basic Sciences, Block JD, Sector III, Salt Lake, Kolkata - 700106, India}

\date{\today}
\begin{abstract}
Two-leg spin-1/2 ladder systems consisting of a ferromagnetic leg and an
antiferromagnetic leg are considered where the spins on the legs interact through  
antiferromagnetic rung couplings $J_1$. These ladders can have two geometrical arrangements either 
zigzag or normal ladder and these systems are frustrated irrespective of their geometry.
This frustration gives rise to incommensurate spin density wave, dimer and spin fluid phases 
in the ground state. The magnetization in the systems decreases linearly
with $J^2_1$, and the systems show an incommensurate phase for $0.0<J_1<1.0$. 
The spin-spin correlation functions in the incommensurate phase follow power 
law decay which is very similar to Heisenberg antiferromagnetic chain in external
magnetic field.  In large $J_1$ limit, the normal ladder behaves like
a collection of singlet dimers, whereas the zigzag ladder behaves as a one
dimensional spin-1/2 antiferromagnetic chain.
\end{abstract}


\end{frontmatter}


\section{\label{sec:intro}Introduction}
The theoretical studies of magnetic spin-1/2 ladder systems have 
been an active area of research because of the existence of 
interesting phases like dimer~\cite{white1994}, spiral 
phase~\cite{mk2015}, different ordered phases~\cite{verkholyak2012},
magnetization plateau~\cite{Honecker2000} etc. 
The spin-1/2 ladder model systems show a rich quantum phase diagrams in various 
interaction coupling limit. 
The Heisenberg antiferromagnetic (HAF) spin-1/2 normal ladder is  
realized in SrCu$_2$O$_3$~\cite{sandvik1995}, (VO)$_2$P$_2$O$_7$ etc.~\cite{dcjhon1987,dagotto96}, 
whereas zigzag ladder, which is considered as the chain 
with nearest and next nearest neighbor interactions, have been realized in (N$_2$H$_5$)CuCl$_3$~\cite{Maeshima2003},  
LiCuSbO$_4$~\cite{dutton_prl}, LiSbVO$_4$~\cite{Mourigal2012}, Li$_2$CuZrO$_4$~\cite{Drechsler2007} etc.
The AF normal ladder system is a spin liquid with a spin gap and 
short range spin correlation. It was conjectured that the spin gap  decreases
smoothly as rung exchange interaction decreases~\cite{dagotto92,white1994,hijii2005} and 
reduces to zero only when rung interaction strength approaches to zero. 
The rung interaction induces the singlet dimer formation between the 
two nearest spin-1/2 on different legs~\cite{dagotto92,dagotto96}. Ladders 
with ferromagnetic legs/rungs and antiferromagnetic rungs/legs are also well studied 
and show interesting phases~\cite{vekua2003,almeida2007,dutton2012,mk_jpcm2013,volkova2012,volkova2016}.
However, the AF zigzag ladder is completely different from normal ladder. The zigzag ladder
 in the weak rung coupling limit $J_1/J_2 < 0.44$ behaves 
like two independent HAF spin-1/2 chains~\cite{mk2016,mk2015,mk2013}, and 
shows gapped spiral phase for $0.44<J_1<2$. It is gapped system  with 
dimer configuration for $2<J_1/J_2<4.148$~\cite{chitra95,mk2016,mk2015,mk2013,ckm69b}.    

In this paper we consider spin ladders which have ferromagnetic (F) spin 
exchange interactions along one of the legs, and antiferromagnetic (AF) 
interactions on the other leg; and spins on these two legs are interacting 
through AF interaction. The focus of this paper is to study some universal theoretical 
aspects such as the existence of exotic phases in the ground state (GS) 
and low-lying excitations in this system. We show that the ferromagnetic-antiferromagnetic (F-AF) ladders pose quasi-long range behavior in
incommensurate regime, and  frustration can be induced even for very small rung coupling limit.

\begin{figure}
\centering
\includegraphics[width=3.3 in]{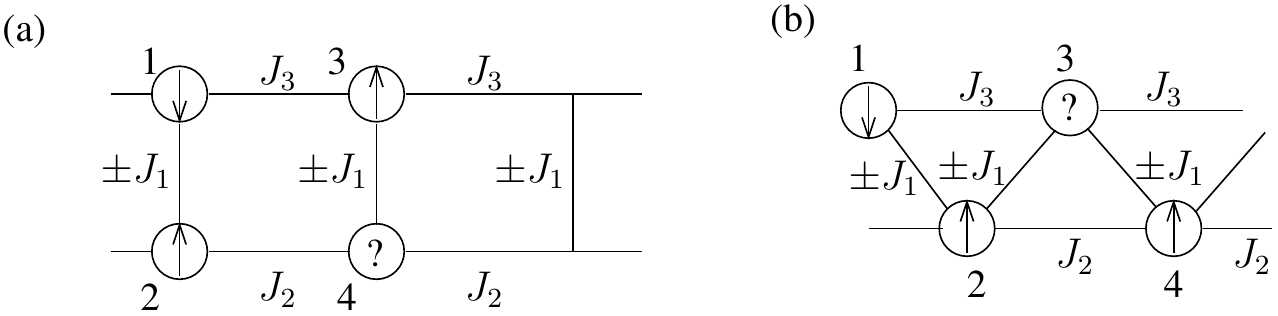}
\caption{\label{fig:figure1}(a) and (b) show the normal and the zigzag arrangements
of the interfaces. The arrows show the spin arrangement and the question mark 
represents the frustrated spin.}
\end{figure}

These two lagged ladders can represent the interface of the two layered magnetic 
spin-1/2 system consisting of an antiferromagnetic and a ferromagnetic layer where the two layers 
interact  with direct or indirect antiferromagnetic exchange. Similar 
 interfaces are studied by Suhl {\it et al.}~\cite{suhl98} and Hong {\it et al.}~\cite{hong98}. We further simplify
the model by considering  only a inter-facial line of spins in the interface of both the layers.  We consider
two possibilities of arrangement of inter-facial spins; first, when spins are directly facing 
each-other as in normal ladder (NL), and second, where spins on one leg is  shifted by half of 
the lattice unit forming a zigzag ladder (ZL).
The spin arrangements of NL and ZL are shown in  the Fig. \ref{fig:figure1}(a) and (b).   
These systems are interesting because both the ladders are frustrated 
irrespective to the nature of rung interactions.


These spin-1/2 NL or ZL  can also give some preliminary information about 
the phases at the interface of bilayer F-AF magnetic 
thin films.   The inter-facial properties of the  F-AF thin film 
materials~\cite{cpbean56,cpbean57} remain a subject of active
research till date. At low temperature (below N\'eel temperature and 
Curie temperature), the spins on both the layers remain ordered. This
leads to an exchange bias at the interface. Many theoretical models 
based on microstructure have been proposed to explain the exchange 
bias field phenomenon~\cite{cpbean56,cpbean57,nogues99,kiwi2001,berkowitz99}
at the interface of these F-AF layers, e.g., discrete micromagnetics
models~\cite{finazzi2004,malozemoff87,koon97,schulthess98,schulthess99,kim2001,nowak2001,sakurai93}, 
continuum micromagnetics models~\cite{mauri87,xi99,xi2000}, and many 
others~\cite{geshev2000,suhl98,schulthess98b}. 
Suhl~\cite{suhl98} considered only the interfacial spins similar to our model and  pointed out, the spins at the antiferromagnetic side of the
interface is in the mean field of the ferromagnetic spins. This happens because
the N\'eel temperature is lower than the Curie temperature, and the spins on the
ferromagnetic side is more robust. 

This paper is divided into four sections. In Section~\ref{sec2}, the model Hamiltonian 
is introduced and the numerical methods are explained. The numerical results 
for both the ladders are given in Section~\ref{sec3} and the effective model 
Hamiltonian is constructed in Section~\ref{sec4}. All the results are discussed and summarized in
Section~\ref{sec5}. 

\section{\label{sec2}Model Hamiltonian and Numerical Method}
We consider a 2-leg ladder (either NL or ZL) of F and AF legs. 
We further consider the  half-filled insulating case where the electrons are 
completely localized, but spins can interact with its nearest neighbors. 
Thus we can write an isotropic Heisenberg spin-1/2 model Hamiltonian for 
the system shown in Figs.~\ref{fig:figure1}~(a) and (b) as
\begin{equation}
H = H_{\mathrm{rung}} + H_{\mathrm{leg}},
\label{eq:ham}
\end{equation}
where 
\begin{eqnarray}
\label{eq:ham_detail}
H_{\mathrm{leg}} &=& \sum_{i=1}^{N/2-1} J_2 \, \vec{S}_{2i} \cdot \vec{S}_{2i+2} 
+ J_3 \, \vec{S}_{2i-1} \cdot \vec{S}_{2i+1}, \nonumber \\
H^{\mathrm{NL}}_{\mathrm{rung}} &=& J_1 \sum_{i=1}^{N/2} \vec{S}_{2i-1} \cdot \vec{S}_{2i}, \\
H^{\mathrm{ZL}}_{\mathrm{rung}} &=& J_1 \sum_{i=1}^{N-1} \vec{S}_{i} \cdot \vec{S}_{i+1}. \nonumber
\end{eqnarray}
Here the rung Hamiltonian for the NL and ZL are written as $H^{\mathrm{NL}}_{\mathrm{rung}}$ 
and  $H^{\mathrm{ZL}}_{\mathrm{rung}}$, respectively. The nearest neighbor AF interaction $J_3$ is 
along the upper leg, and nearest neighbor F interaction $J_2$ is along the 
lower leg. $J_1$ is interaction along the rung of the systems as shown in 
Fig.~\ref{fig:figure1}(a) and Fig.~\ref{fig:figure1}(b) for the NL and the ZL, respectively. 
The interactions along legs are set to $J_2= -1$ and $J_3=1$, 
however, rung interaction  ($J_1=\alpha$) is a variable quantity. To understand 
the GS properties of these systems as a function of $\alpha$,
we solve the Hamiltonian in Eq.~(\ref{eq:ham}) numerically.

We use the exact diagonalization (ED) method for small systems and Density matrix renormalization 
group (DMRG) method to handle the large degrees of freedom 
for large systems. The DMRG is based on the systematic 
truncation of irrelevant degrees of freedom at every step of growth of the chain~\cite{white-prl92,karen2006,schollwock2005}. We have 
used recently developed DMRG method where 
four new sites are added at every DMRG steps~\cite{mk2010b}. We have also used
the recently developed DMRG for periodic boundary condition (PBC) when the 
system is under PBC~\cite{ddpbc2016}. The eigenvectors 
corresponding to $m$ largest eigenvalues of the density matrix of the system in 
the GS of Hamiltonian in Eq.~(\ref{eq:ham}) are kept to construct the effective 
density matrix. We have kept $m$ up to 500 to keep the truncation error 
less than $10^{-10}$. We have used system sizes up to $N=200$ to minimize 
the finite size effect.  

\section{\label{sec3}Numerical Results}
In this section, we analyze the GS of both the ZL and the NL 
for various rung interaction ($\alpha$) limits. Here we consider 
only the antiferromagnetic inter-chain interaction i.e., $\alpha > 0$.
In the small $\alpha \,(\ll 1)$ limit, the NL and the ZL behave like decoupled chains. 
In this phase the F leg remains in ferromagnetic state, whereas other leg possesses 
antiferromagnetic arrangement of spins. However, in thermodynamic limit the 
decoupled phase exists only for  $\alpha \sim 0$. On further increase in $\alpha$, the competition 
between the F and the AF interactions forces the F leg to reduce its total magnetization.  
There is an incommensurate spin density wave (SDW) phase for parameter space 
$0.07 < \alpha < 1.14$ in the NL and $0.04 < \alpha < 1.06$ for the ZL with $N=200$ spins. In 
thermodynamic limit, the lower limit of $\alpha$ value for SDW phase tends to zero. In the 
large $\alpha$ limit of the NL, the two nearest neighbor spins from different legs form 
a singlet dimer. However, the ZL behaves like a single spin-1/2 chain of $N$ spins where each leg 
contains $N/2$ spins.  To verify the above phases, total magnetization $\langle M \rangle$ 
in GS, correlation functions $C(r)$, and the spin densities $\rho_r$ for 
both the systems are analyzed.  

\subsection{\label{subs1}Magnetization}
\begin{figure}[t]
\centering
\includegraphics[width=3.0 in]{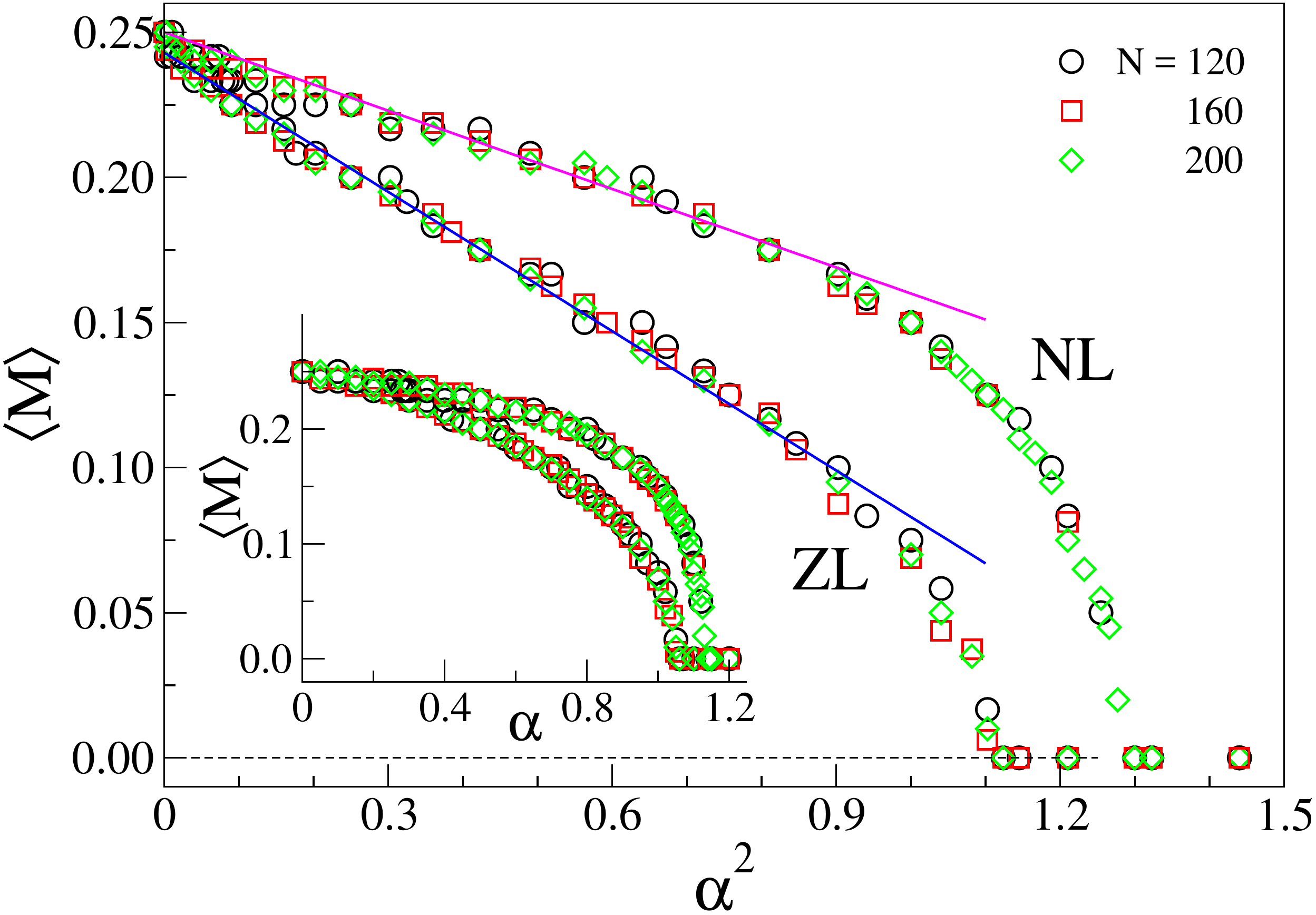}
\caption{\label{fig:szjump} The ground state magnetization $\langle M \rangle$ with 
inter chain coupling interaction $\alpha$ for both the NL and the ZL. The main 
figure plots $\langle M \rangle$ vs. $\alpha^2$; the straight line fits for $\alpha < 0.9$
i.e., $\alpha^2 < 0.8$ reveals that $\langle M \rangle \propto \alpha^2$ in this regime.
The inset shows the $\langle M \rangle$ - $\alpha$ curve.}
\end{figure}

For small inter-chain antiferromagnetic coupling ($\alpha \ll 1$), two legs of the ladder behave as
decoupled chains, and the system has its ground state magnetization 
$\langle M \rangle =\frac{\sum_{i=1}^N S_i^z}{N} = \frac{1}{4}$. All of the 
magnetization contribution comes from the F leg. The magnetization 
$\langle M \rangle$ decreases continuously with $\alpha$, and $\langle M \rangle$ goes finally 
to zero for large $\alpha$. The $\langle M \rangle$ as a function of $\alpha^2$ is shown in 
Fig.~\ref{fig:szjump} (main) for three system sizes $N =120,160$, and 200 of both the NL and the ZL 
systems, and the inset show the $\langle M \rangle$ -- $\alpha$ curve.  We notice there 
are step like behavior in $\langle M \rangle-\alpha$ plot  in finite system, 
but width of steps decreases with system size $N$. However,  $\langle M \rangle - \alpha$ curve
should be continuous in the thermodynamic limit. We notice that the NL shows slower change in $\langle M \rangle - \alpha$
as compared to the ZL. 

The transition from incommensurate SDW phase to spin fluid phase 
at $\alpha=\alpha_c=1.06$ is relatively faster in the ZL compared to the transition from incommensurate 
SDW to dimer phase at $\alpha=\alpha_c=1.14$ in the NL, and the $\langle M \rangle$ vanishes at the 
transition point  $\alpha=\alpha_c$. In fact our analytical perturbation calculation for the 
NL in Sec.~\ref{sec4} also suggests that the contribution from interaction along the legs are zero 
at large $\alpha$ limit. In this limit the NL system of $N$ spins  behaves as a collection of 
$\frac{N}{2}$ number of independent singlet dimers. The continuous variation in the ZL near 
the transition point can be attributed to delocalized nature of the system. In this limit 
$J_1$ dominates, and this system behaves like a HAF spin-1/2 chain with weak and alternate AF 
and F next nearest neighbor interaction. The ferromagnetic interaction $J_2$ stabilize the 
AF arrangement of spins, whereas AF interaction $J_3$  frustrates the system.

To understand the spin arrangement and correlation between the spins, spin correlations and spin densities are studied.

\subsection{\label{subs2}Spin-spin correlations}
Longitudinal spin-spin correlations are defined as 
\begin{equation}
C(r) = \langle S_i^z S_{i+r}^z \rangle 
\label{eq:cr}
\end{equation}
where $S^z_i$ and $S_{i+r}^z$
are the z-component of spin operators at reference site $i$ and at a distance $r$ 
from the reference site $i$, respectively. Our reference site is at the AF leg in this 
subsection. We have also defined spin density fluctuation as 
\begin{equation}
C^{\mathrm{F}}(r)=\langle S_i^z S_{i+r}^z \rangle - \langle S_i^z \rangle \langle S_{i+r}^z \rangle.
\label{eq:cfr}
\end{equation}
\begin{figure}
\centering
\includegraphics[width=3.0 in]{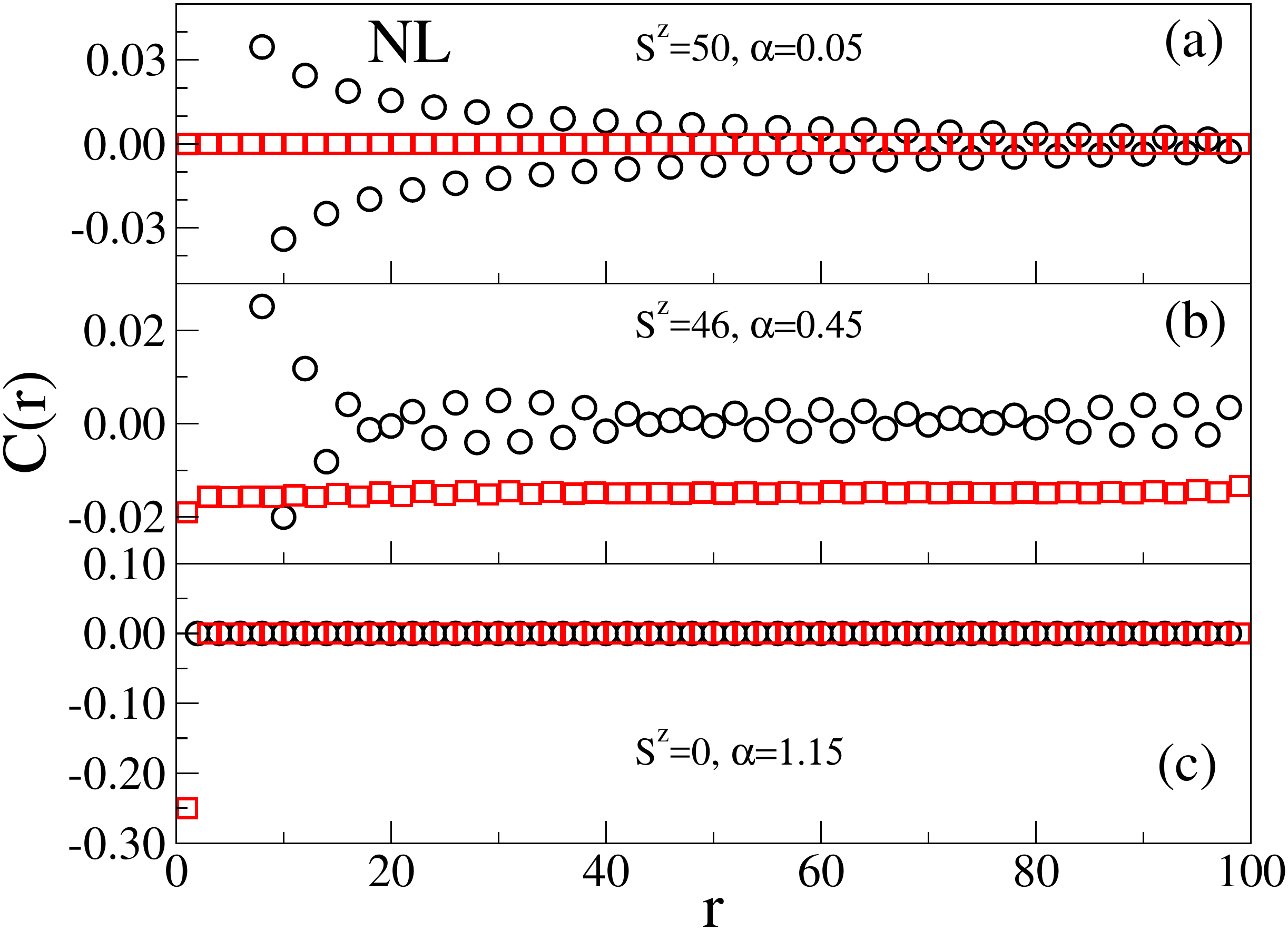}
\includegraphics[width=3.0 in]{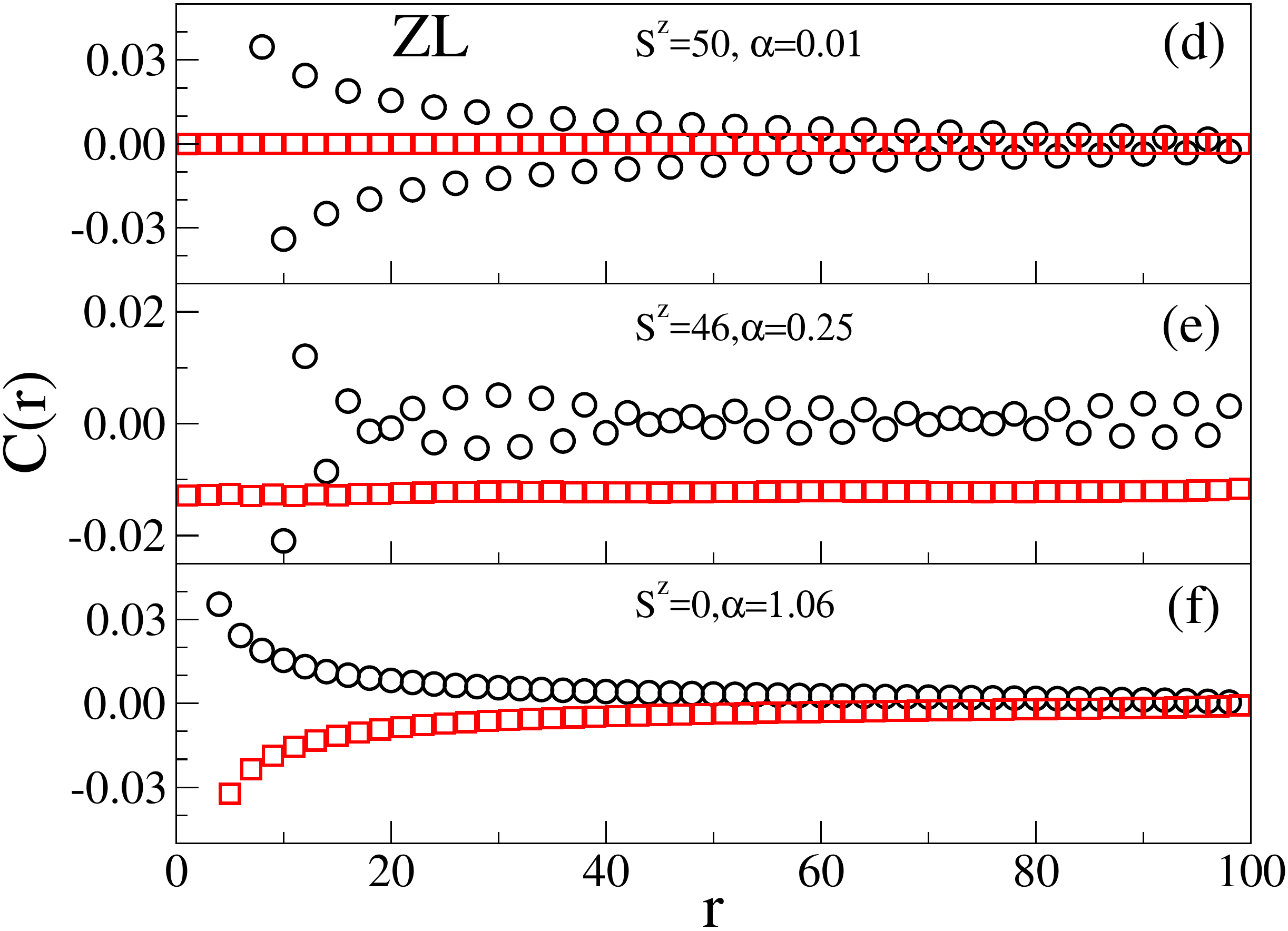}
\caption{\label{fig:ful_corl} The longitudinal spin-spin correlations $C(r)$ 
are plotted for N=200 considering the reference spin on the AF leg. Here circles 
represent $C(r)$ with the spins of the AF leg whereas squares represent $C(r)$ 
with the spins of the  F leg. Different values of $\alpha$ are chosen 
to show the (a) decoupled phase, (b) incommensurate SDW phase, (c) dimer phase 
for $\alpha=0.05$, 0.45, 1.15 respectively in the NL, 
and (d) decoupled phase, (e) incommensurate SDW phase, (f) spin-fluid phase 
for $\alpha=0.01$, 0.25, 1.06 respectively in the  ZL.}
\end{figure}
We find three types of correlations for both the ladder systems in different parameter 
regimes as shown in Fig.~\ref{fig:ful_corl}. Black circles represent correlations 
with spins located on AF leg, whereas squares represent the correlations with the 
spins on F leg. $C(r)$ for three different phases are shown in 
Figs.~\ref{fig:ful_corl}(a)--(c) for the NL and Figs.~\ref{fig:ful_corl}(d)--(f) for the ZL with N=200. 
For $\alpha=0.05$, the spins on different legs are uncorrelated, and  the spins on the AF 
leg show quasi-long-range order, as shown in Fig.~\ref{fig:ful_corl}(a). The similar 
behavior is found in the ZL for $\alpha = 0.01$ as shown in Fig.~\ref{fig:ful_corl}(d). 
The incommensurate phase in the NL is observed for $0.07 < \alpha < 1.14$ and 
in the ZL for $0.04 < \alpha < 1.06$. We choose $\alpha = 0.45$ for the NL and $\alpha = 0.25$ 
for the ZL  to make sure that  we have same $S^z$ value in both types of ladder. At 
large distance $r$, the value of $C(r)$ for both the  NL and the ZL is finite 
as shown in Fig.~\ref{fig:ful_corl}(b) for NL and Fig.~\ref{fig:ful_corl}(e) for ZL. However, $C^{\mathrm{F}}(r)$ decays 
algebraically for the AF and the F leg separately where we consider the reference 
spin on the AF and the F leg, respectively. 

At large $\alpha$ limit ($\alpha \ge 1.16$), 
the behavior of the two ladders become completely different. In NL the  $C(r)$ have 
non-zero value only up to $r = 1$, as shown in Fig.~\ref{fig:ful_corl}(c). 
In the large $\alpha \,(\ge 1.06)$ limit the ZL behaves like a single antiferromagnetic 
Heisenberg chain and the $C(r)$ decays following a power law ($\propto r^{-\gamma}$) 
where spins from AF and F leg situated alternatively with distance $r$. The $C(r)$ in 
this regime is shown in Fig.~\ref{fig:ful_corl}(f) for $\alpha = 1.06$.
\begin{figure}
\centering
\includegraphics[width=3.0 in]{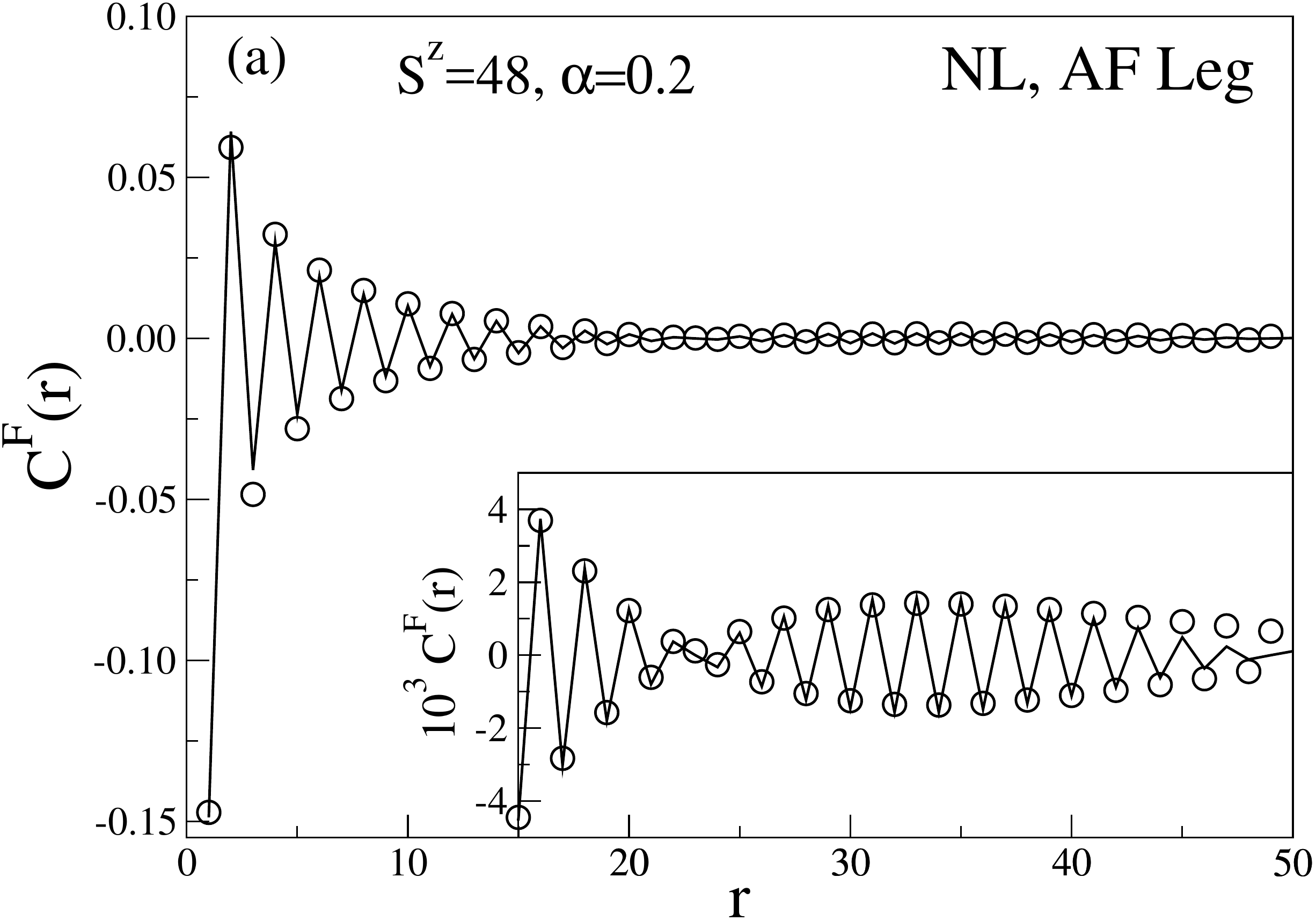}
\includegraphics[width=3.0 in]{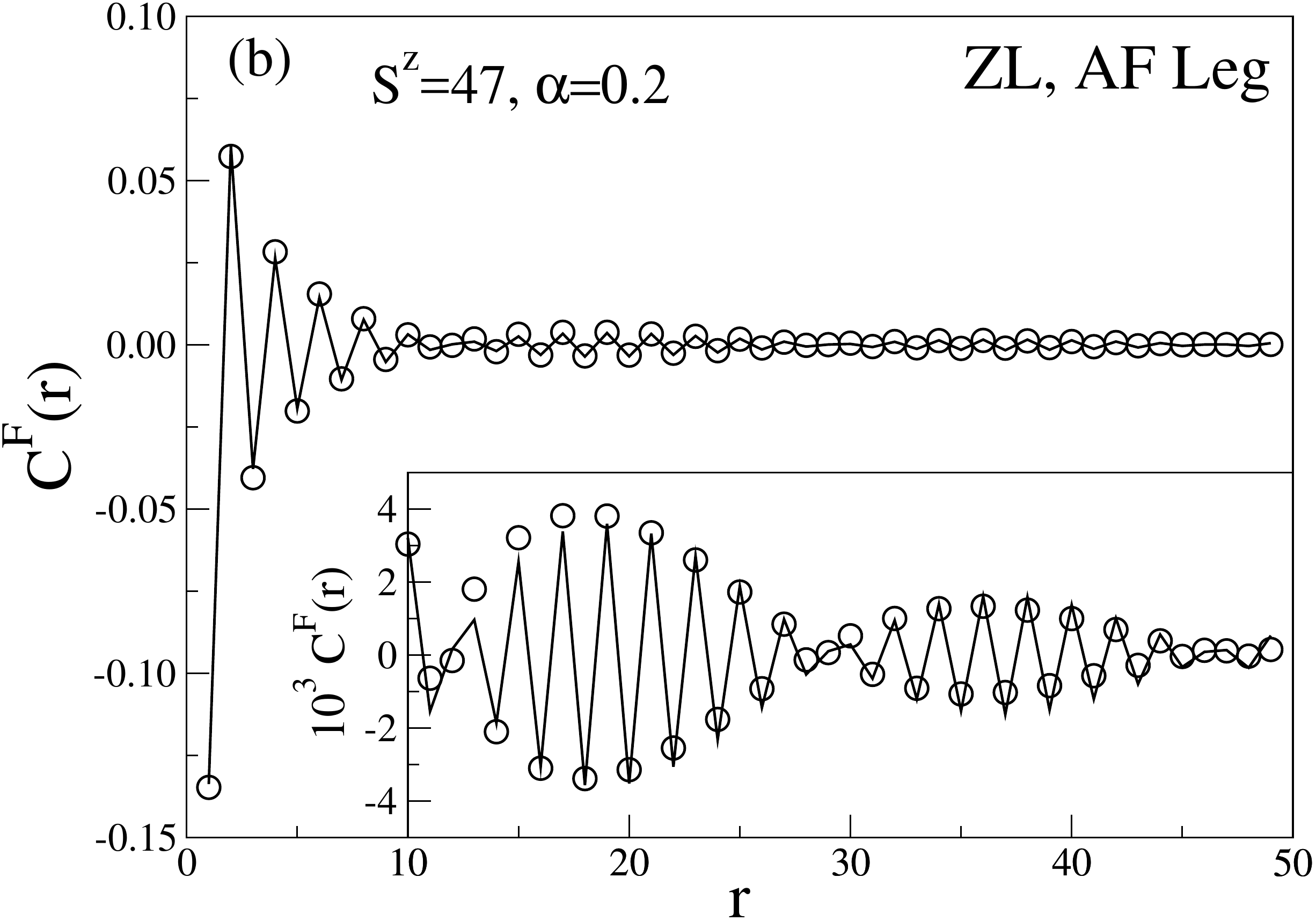} 
\caption{\label{fig:corl_inset}Spin density fluctuations $C^{F}(r)$ for the 
spins on the  AF leg for (a) NL and (b) ZL. The reference spin is on the AF leg.
The points are the ground state $C^{\mathrm{F}}(r)$ calculated using DMRG for $N=200$, 
$\alpha = 0.2$ for both the ladders. Calculated $C^{\mathrm{F}}(r)$ are fitted using 
Eq.~(\ref{eq:fit}) and the solid lines represent the fitted curves. The insets 
are the zoomed $C^{\mathrm{F}}(r)$ plot.}
\end{figure}

The incommensurate SDW phases for the NL and the  ZL are similar to 
that in the HAF spin-1/2 chain in a magnetic field; therefore,
$C(r)$ in the F and the AF legs are  analyzed separately for both the  NL and the ZL. 
$C^{\mathrm{F}}(r)$ for the F leg in both the systems are vanishingly small. To 
understand it better, we plot  $C^{\mathrm{F}}(r)$ for the AF leg  in 
Figs.~\ref{fig:corl_inset}(a) and (b) for the NL and the ZL, respectively, 
for same $\alpha \,(=0.2)$ and $N=200$. 
We find that $C^{\mathrm{F}}(r)$ in AF leg for both the systems follow the relation 
\begin{equation}
C^{\mathrm{F}}(r) \propto (-1)^r \, r^{-\gamma} \, \sin\left(\frac{\pi (r+c)}{\beta}\right)
\label{eq:fit}
\end{equation}
Here $\beta$ is proportional to wavelength of the SDW, $\gamma$ represents 
the power law coefficient, and $\frac{\pi c}{\beta}$ is a phase shift. 
In Fig.~\ref{fig:corl_inset}(a) and Fig.~\ref{fig:corl_inset}(b) solid lines are the fitted curves
where the symbols indicate numerically calculated values of $C^{\mathrm{F}}(r)$. 
$\beta$ depends on the value of $\alpha$. In the main Fig.~\ref{fig:corl_inset}(a), 
$C^{\mathrm{F}}(r)$ for $\alpha = 0.2$ for the NL is shown, and the values are fitted using Eq.~(\ref{eq:fit}) 
with the values $\beta=25$ and $\gamma=1.5$.  
Fig.~\ref{fig:corl_inset}(b) shows the same for the ZL and the fitted 
parameter values are  $\beta=17$ and $\gamma=1.25$. The insets of Fig.~\ref{fig:corl_inset}(a)
and Fig.~\ref{fig:corl_inset}(b) are the zoomed $C^{\mathrm{F}}(r)$ and these show that Eq.~(\ref{eq:fit}) fits very well
even for large distances. The variation of $\beta$ with $\alpha$ is discussed in the 
subsections~\ref{subs3}~and~\ref{subs4}.  The $C^{\mathrm{F}}(r)$ on the F leg have values of order of $10^{-5}-10^{-6}$; therefore,
it is difficult to exactly fit the $C^{\mathrm{F}}(r)$ values on the F leg.

\subsection{\label{subs3}Spin density}
\begin{figure*}
\centering
\includegraphics[width=5.2 in]{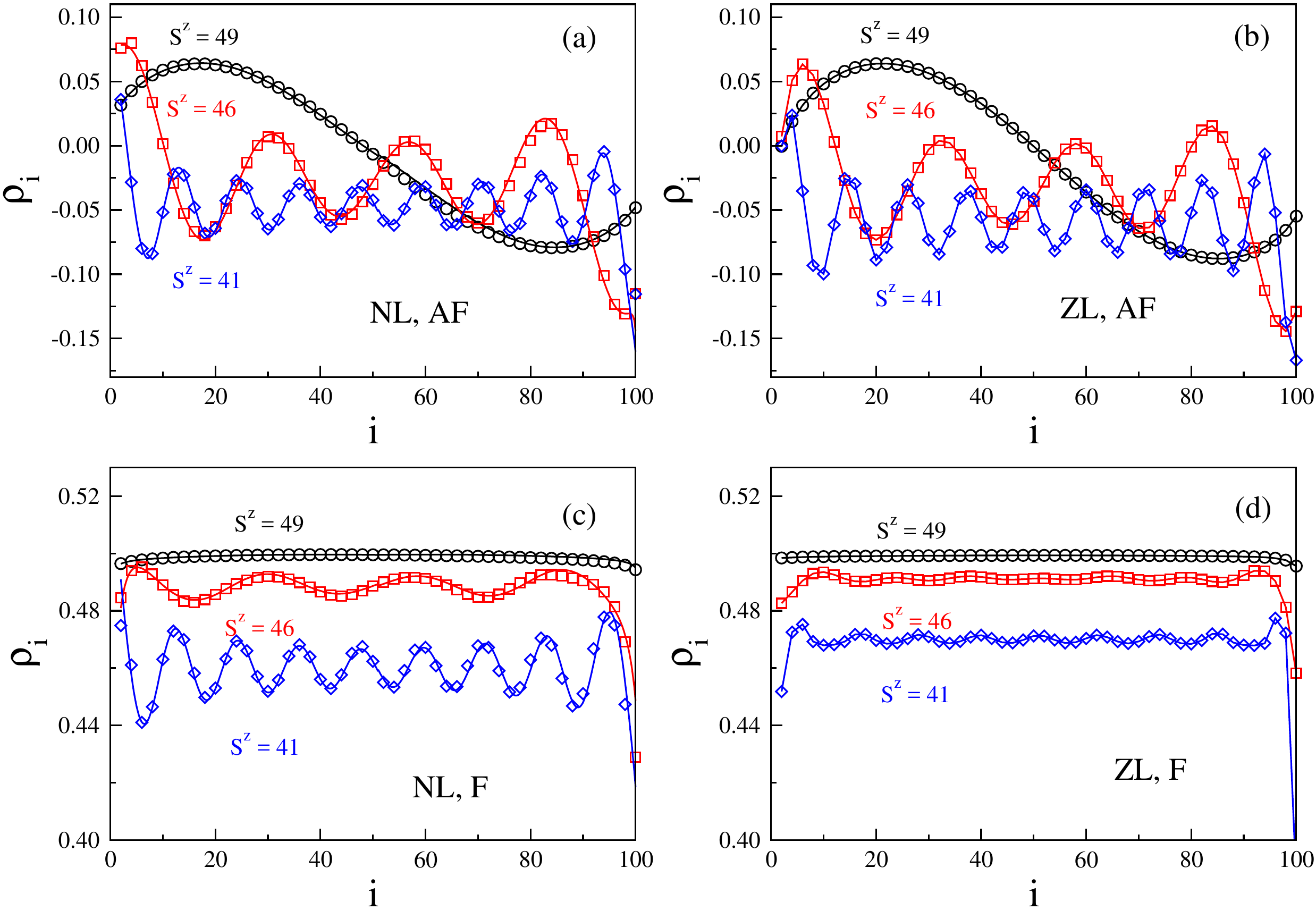}
\caption{\label{fig:spden} Spin densities for alternate sites on the AF 
leg and the F leg of the NL and the ZL for $S^z = 49$, 46 and 41. 
(a) and (c) depict the spin densities on AF leg and F leg of NL
for $\alpha = 0.1$, 0.45, 0.75 respectively; (b) and (d) show the spin 
densities on the AF leg and F leg for ZL for $\alpha = 0.1$, 0.25, 0.45 
respectively.}
\end{figure*}

The distribution of the spin density on different legs is important, especially in 
the higher magnetic states. The spin densities on the odd (even) sites  correspond
to the spin densities on the AF (F) leg. In 
Figs.~\ref{fig:spden}(a) and (b), the spin densities of alternate sites in the AF leg are shown for the 
NL and the ZL respectively.  
For small $\alpha$, three $S^z$ sectors are considered at different
$\alpha$. The incommensurate spin density ($\rho_i$) for  $S^z = 49$, 46, and 41 for 
$\alpha = 0.1, 0.45$, and $0.75$ in the AF leg of the NL are shown in 
Fig.~\ref{fig:spden}(a). $\rho_i$ for same values of $S^z$ for $\alpha = 0.1, 0.25$, 
and $0.45$ in the AF leg of the ZL system are shown in Fig.~\ref{fig:spden}(b). 
In these systems the incommensurate SDW phase shows the similar behavior for a given 
$S^z$ except at the boundary of the system. $\rho_i$ in the F leg for both the NL 
and the ZL also show incommensurate SDW as shown in Figs.~\ref{fig:spden}(c) and (d), respectively. 
However, the incommensurate SDW is more prominent in smaller $S^z$. 

\begin{figure}
\centering
\includegraphics[width=3.0 in]{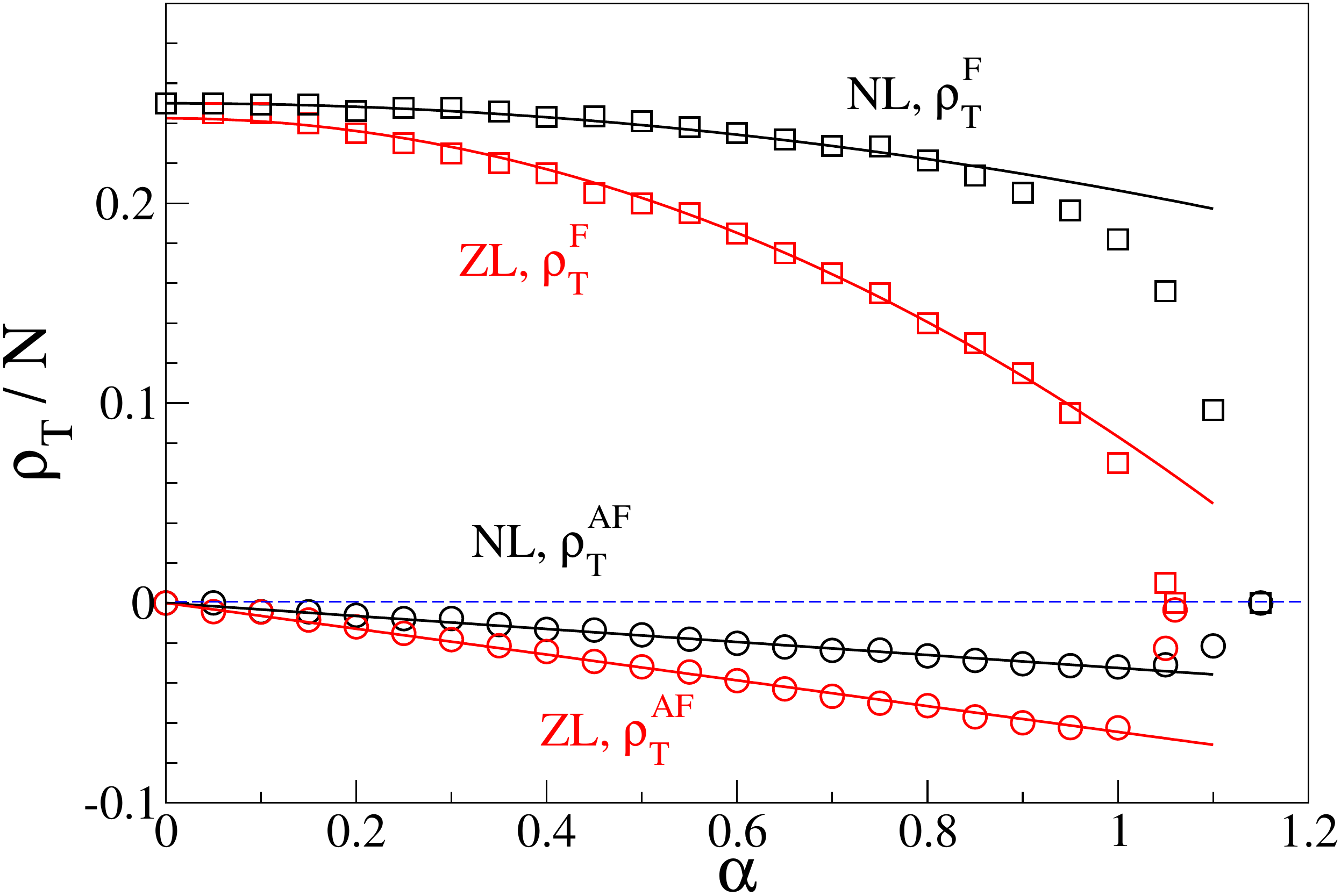}
\caption{\label{fig:sztot}Total spin density on the AF leg and the F leg for 
both NL and ZL for different values of $\alpha$. For $\alpha < 0.9$ in the F leg 
$\rho_T^F \propto \alpha^2$ while in the AF leg $\rho_T^{AF} \propto \alpha$.}
\end{figure}

The edge spin densities of the F leg in the NL is much smaller than the 
ZL system. The spin density modulation in the F leg is very small for $\alpha \to 0$;
however, the amplitude of the modulations increase with $\alpha$. For a given $S^z$ 
incommensurate SDW in ZL has smaller amplitude than that in NL, but both have
similar $\beta$. For large $\alpha < \alpha_c$, the periodicity of the AF and the F legs are 
the same. The total spin density, on  each leg of 
the system, $\rho_{T}$ is shown as a function of $\alpha$ in Fig.~\ref{fig:sztot}. 
The circle and square of same color  represent $\rho_T$ for the AF and the F leg, respectively, 
however black and red symbols represent the NL and the ZL, respectively.  The   
$\rho_T$ value in the AF leg is zero in decoupled limit, and it varies 
linearly with $\alpha$ with negative slope in both the systems for $\alpha < 0.9$. 
However, for $\alpha > 0.9$, $\rho_T$ increases rapidly, and goes to zero at 
the transition point ($\alpha_{\mathrm{c}}$). In the F leg of both systems, $\rho_{T}$ decreases 
monotonically with $\alpha^2$ for $\alpha < 0.9$; $\rho_{T}$ decreases rapidly to zero near the transition point 
$\alpha_{\mathrm{c}}=1.14$ ($1.06$) for the NL (the ZL).

If we ignore some points near the edges, the spin densities in Fig.~\ref{fig:spden} 
can be fitted with the equation which is proportional to $\sin(\frac{\pi (r+c)}{\beta})$ 
part in Eq.~(\ref{eq:fit}). For the same value of $\alpha$, $\rho_i$ and $C^{\mathrm{F}}(r)$ have same 
$\beta$ for a particular system. The lowest density amplitudes at the edges is due to the 
boundary effect. The AF leg has highest density at the edge and induces highest 
fluctuation in the F leg. The incommensurate SDW has well defined pitch~angle. 

\subsection{\label{subs4}Pitch angle} 
\begin{figure}
\centering
\includegraphics[width=3.0 in]{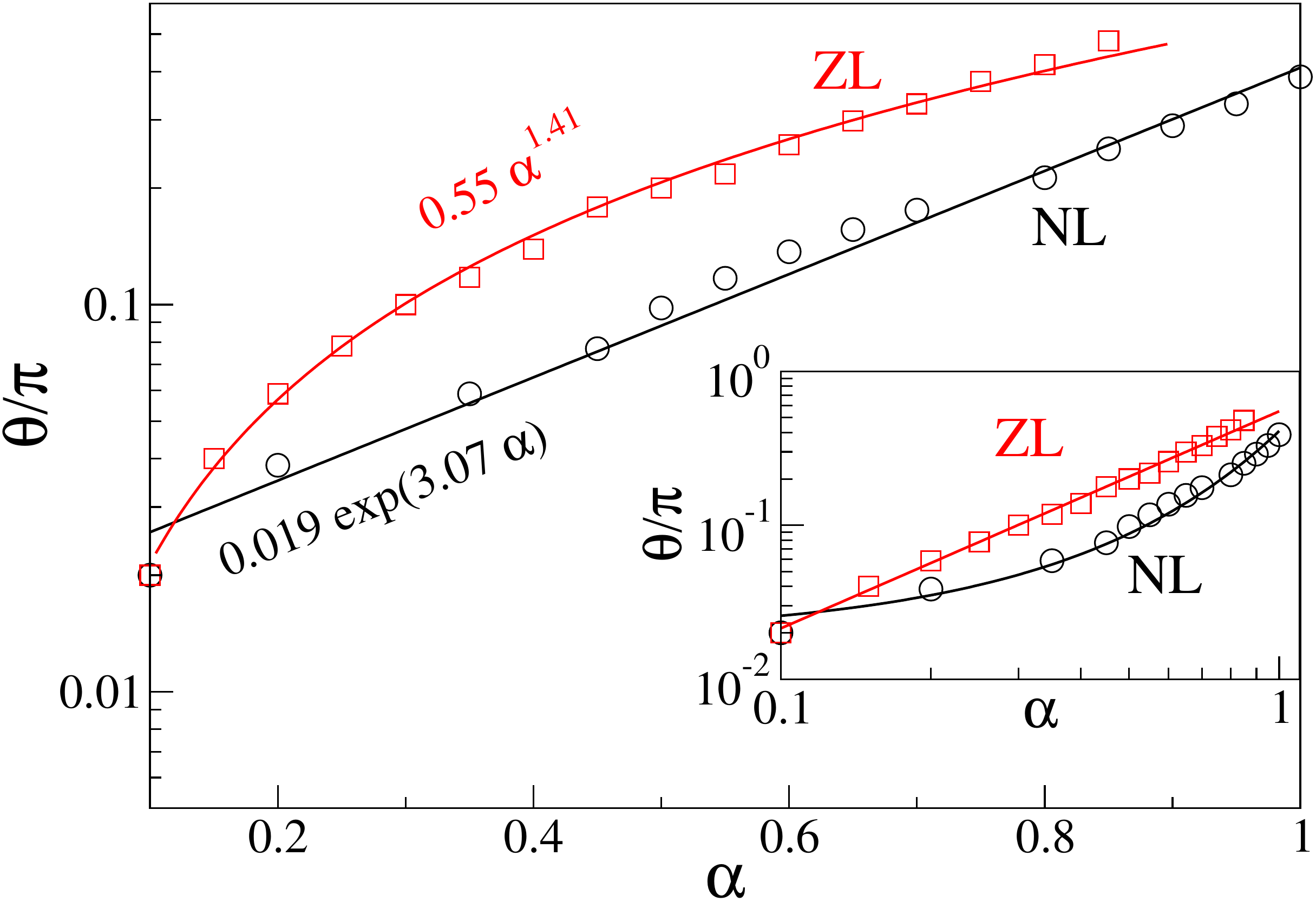}
\caption{\label{fig:pitch}The pitch angle ($\theta$) for different values of $\alpha$
in both the NL and the ZL in log-linear scale (main figure) and in log-log scale (inset).
The solid lines are the exponential fit (for NL) and the power law fit (for ZL). The
actual fitted formulae are given on the plot near the curves.}
\end{figure}

We notice that accurate calculation of  the pitch angle ($\theta$) from $C^{\mathrm{F}}(r)$ 
becomes extremely difficult because of power law nature of  $C^{\mathrm{F}}(r)$.  However, 
$\theta$ can be directly calculated from the spin density calculations. Now let 
us consider the length of the AF chain is $l$ for which the total  angle change 
between $i^{\mathrm{th}}$ and $i+l^{\mathrm{th}}$ spin is $2\pi$. The length $l$ 
is the wavelength of the incommensurate SDW. Therefore $\theta$ can be defined as $2\pi/l$. 
The pitch angle $\theta$ in the AF leg as a function of $\alpha$ is plotted in 
Fig.~\ref{fig:pitch} for the NL with black circle and for the ZL  with red  
squares. The main figure shows the log-linear plot. The pitch angle 
for the NL follows exponential  decay as shown in the main Fig.~\ref{fig:pitch}. 
The line represents the  fitted curve with $0.019 \exp(3.07 \alpha)$. The convention of 
symbols in the inset is the same as in the main figure. The inset of Fig.~\ref{fig:pitch} 
shows the log-log plot of $\theta /\pi \mathrm{ vs. } \alpha$, which is fitted with 
$0.55\, \alpha^{1.41}$ for the ZL. 

\subsection{\label{subs5}Large $\alpha$ limit}  
In the large $\alpha >1.14$ limit, the NL behaves as dimer of two nearest neighbor 
spins at different leg of ladder. The $C(r)$ for $\alpha=1.15$ is very short ranged 
and is non-zero only for nearest neighbor spins as shown in Fig.~\ref{fig:ful_corl}(c). The GS energy  
is exactly equal to $-\frac{3}{8} N \alpha$.  The AF 2-leg ladder has 
a short range order and has finite lowest energy gap (spin gap)  $\Delta = -3J_1/4$ in the perfect dimer limit.
This energy is equivalent to breaking a singlet bond. The short range correlation in NL 
and gaped excitation are explained in the analytical Section~\ref{sec4}.

In this limit the ZL system behaves as a single chain of $N$ spins 
with nearest  neighbor exchange interaction $J_1$. 
It is well known that the GS of spin-1/2 HAF chain is  a spin fluid state and 
this phase can be characterized by the algebraic decay of spin-spin correlations 
and the gapless energy excitation. The transition point from the incommensurate SDW phase to 
the spin fluid in the  ZL is at $\alpha \approx 1.06$.

\section{\label{sec4}Effective Hamiltonian for NL}
In large inter-chain coupling limit the NL shows dimer phase as already mentioned
in Section~\ref{subs5}. To understand the dimer phase we treat the NL analytically in this 
section. Our aim is to find an effective Hamiltonian for the NL in the strong coupling 
limit i.e., for $\alpha >> 1$. There is a very sharp critical $\alpha$
for NL and the system has singlet GS for $\alpha > \alpha_c$. 

In the strong rung coupling limit the system can be approximated 
as a collection of $N/2$ rungs. The Hamiltonian then becomes 
$H_{\mathrm{rung}}^{\mathrm{NL}}$ (see Eq.~(\ref{eq:ham_detail})).
The GS of this Hamiltonian is  $2^{N/2}$ fold degenerate. Each of $N/2$ rungs
can be either in the state $|S_0 \rangle$ or $|T_1 \rangle $ with energies
$E(S_0) = -3J_1 / 4$ and $E(T_1) = J_1 / 4$.
$H_{\mathrm{leg}}$  lifts the degeneracy and leads to an effective Hamiltonian that can be derived 
by standard many-body perturbation theory~\cite{fulde95}. Following the same procedure 
mentioned in \cite{mila98}, we can write the spin operators in terms of the 
pseudo-spin-1/2 operators. Let us introduce pseudo spin-1/2 operators 
$\mathbf{\tau}_i$ to be acted on the states $|S_0 \rangle_i$ and $|T_1 \rangle_i$ of rung $i$ following
\begin{eqnarray}
\tau_i^z |S_0 \rangle_i =-\frac{1}{2} |S_0 \rangle_i & \qquad  & \tau_i^z |T_1 \rangle_i =\frac{1}{2} |T_1 \rangle_i \nonumber \\
\tau_i^+ |S_0 \rangle_i = |T_1 \rangle_i & \qquad & \tau_i^+ |T_1\rangle_i = 0 \\
\tau_i^- |S_0 \rangle_i = 0 & \qquad & \tau_i^- |T_1\rangle_i = |S_0 \rangle_i \nonumber
\end{eqnarray}
One can express the original operators in Eq.~(\ref{eq:ham_detail}) in terms of the
pseudo-spin operators. This can be done by inspection and are given by:
\begin{eqnarray}
S_{2i-1}^+ = -\frac{1}{\sqrt{2}} \tau_i^+ & \qquad & S_{2i+1}^+ = \frac{1}{\sqrt{2}} \tau_i^+ \nonumber \\
S_{2i-1}^- = -\frac{1}{\sqrt{2}} \tau_i^- & \qquad & S_{2i+1}^- = \frac{1}{\sqrt{2}} \tau_i^- \\
S_{2i-1}^z = \frac{1}{2} \left(\tau_i^z + \frac{1}{2} \right) & \qquad & S_{2i+1}^z = \frac{1}{2} \left(\tau_i^z + \frac{1}{2} \right) \nonumber
\end{eqnarray}
Substituting these expressions into the Hamiltonian we get the effective Hamiltonian as:
\begin{eqnarray}
H &=& H_0 +J_{xy}^{\mathrm{eff}} \sum_{i=1}^{N/2} \frac{1}{2} \left(\tau_i^+ \tau_{i+1}^- + \tau_i^- \tau_{i+1}^+ \right) \nonumber \\
  & &  \qquad  + J_z^{\mathrm{eff}} \sum_{i=1}^{N/2} \tau_i^z \tau_{i+1}^z + h_{\mathrm{eff}} \sum_{i=1}^{N/2} \tau_i^z + C_0
\end{eqnarray}
where $J_{xy}^{\mathrm{eff}} = \frac{J_1 + J_2}{2}$, $J_z^{\mathrm{eff}} =  \frac{J_1 + J_2}{4}$, $h_{\mathrm{eff}} =  \frac{J_1 + J_2}{4}$,
and $C_0 = \frac{J_1 + J_2}{4}$.
For the AF-F ladder  $J_1 = -J_2 = J$, therefore at strong coupling limit $H = H_0$. 
The critical rung interactions are given in the Appendix.

\section{\label{sec5}Discussion and Conclusions}
In this paper We studied F-AF ladder models with an antiferromagnetic leg
and a ferromagnetic leg coupled through antiferromagnetic rungs.
Two types (NL and the ZL) of ladders are considered (see Fig.~\ref{fig:figure1}).  
It is clear from Figs.~\ref{fig:figure1}(a) and (b)
that irrespective of the structure both NL and ZL are 
frustrated in nature. Our calculation suggests that both NL and ZL show 
incommensurate SDW phase for $0 < \alpha < \alpha_c$ in the thermodynamic limit. 

We show that  the ZL exhibits some similarities with the NL in the low $\alpha$ 
limit, but have remarkable differences in the large $\alpha$ limit. In small 
and intermediate $\alpha$ limit both ladders exhibit incommensurate SDWs. In 
large $\alpha$ limit the ZL behaves like single HAF chain and the NL shows an 
exact dimer phase. The pitch angle in the incommensurate SDW phase of the ZL 
show algebraic variation, whereas it changes  exponentially in the NL. The spin 
density fluctuation $C^{\mathrm{F}}(r)$ follows the power law in both legs for 
both type of ladders as shown in Fig.~\ref{fig:corl_inset}; this behavior is 
similar as partially magnetized HAF chain. Most of HAF ladder~\cite{dagotto92,dagotto96}
or frustrated ladder~\cite{white96, mk2013,chitra95} shows exponential 
behavior of correlation function in the spiral phase.
The critical value $\alpha_c$ for incommensurate SDW to singlet dimer 
transition in NL is almost independent of system size, and can be 
explained by analytical calculation of perturbation theory of this system. 
The finite size effect on $\alpha_c$ in ZL is also weak. \\

For $\alpha < 0.9$ in both the ladders $\langle M \rangle \propto \alpha^2$ as 
shown in Fig.~\ref{fig:szjump}. $\langle M \rangle$ decreases rapidly to zero 
for  $\alpha > 0.9$. Interestingly for both the ladders in $\alpha < 0.9$ regime, 
the total spin density on the F leg $\rho^{F}_T \propto \alpha^2$ whereas on the 
AF leg $\rho^{F}_T \propto \alpha$ for $\alpha < 0.9$ as shown in Fig.~\ref{fig:sztot}.
The spin density $\rho^{AF}_T$ at AF-leg of both the systems is always negative, 
and $\rho^{AF}_T$ for the ZL have higher magnitude than the NL contrary to the $\rho^{F}_T$.  


In the mean field limit this model can be approximated as a partially magnetized
HAF chain, at least in small coupling limit. Here the F-leg act as a uniform 
external magnetic field on the AF-leg. Using the Heisenberg Hamiltonian with 
anisotropy constant,  Suhl and Schuller explained the effective bias field 
$H_{\mathrm{eff}} \propto J^2_c$ in Eq.~(12) of~\cite{suhl98}, where $J_c$ is 
the exchange interaction strength between two layers. However as shown in 
Fig.~\ref{fig:szjump}, $\langle M \rangle \propto \alpha^2$ in both the systems 
for $\alpha < 0.9$. Assuming the magnetization in these systems is proportional 
to field ($h$), we obtain $h \propto \alpha^2$. T. M. Hong suggested in~\cite{hong98} 
that in low temperature limit $h \propto J_c$. Our calculation agrees very well 
with the calculations in \cite{suhl98}   

In conclusion, we consider F-AF two-legged spin-1/2 ladders with antiferromagnetic
rungs. In the finite $\alpha < 1.0 $ regime we 
notice the incommensurate SDW phase in both the ladders. We also notice that the NL behaves like a 
collection of independent dimers for $\alpha >1.14$, whereas the ZL behaves like 
a single spin-1/2 chain for $\alpha >1.06$. The magnetization 
on the F leg varies as $J^2_1$, whereas it varies linearly 
with $J_1$ on AF leg. 


\section*{Acknowledgement}
MK thanks DST for Ramanujan fellowship and computation facility provided under the 
DST project SNB/MK/14-15/137. We thank Z. G. Soos for his valuable comments.

\appendix
\section{Critical rung interaction}
\sloppy
According to the numerical analysis we find a sharp transition from the $S_G>0$
to $S_G=0$ at some critical value of $\alpha$ for both the ladders. This
critical value is independent of the system size. 
Let us consider a toy model of four spins as shown in Fig.~\ref{fig:figure1}(a).
The eigenvalues in the $S^z = 0$ sector are $\frac{1}{4} (-2J_1 + J_2 + J_3)$, 
$\frac{1}{4} (2J_1 + J_2 + J_3)$, $-\frac{1}{4}\left(J_2+J_3+ 2\sqrt{(J_2-J_3)^2 + J_1^2}\right)$, 
$-\frac{1}{4}\left(J_2+J_3- 2\sqrt{(J_2-J_3)^2 + J_1^2}\right)$,
$-\frac{1}{4}\left(2 J_1+J_2+J_3+2\sqrt{(-J_1+J_2+J_3)^2+3J_1^2}\right)$, 
$-\frac{1}{4}\left(2 J_1+J_2+J_3-2\sqrt{(-J_1+J_2+J_3)^2+3J_1^2}\right)$ and the eigenvalues
in the $S^z = 1$ sector are $\frac{1}{4} (-2J_1 + J_2 + J_3)$,
$\frac{1}{4} (2J_1 + J_2 + J_3)$, $-\frac{1}{4}\left(J_2+J_3+ 2\sqrt{(J_2-J_3)^2 + J_1^2}\right)$,
$-\frac{1}{4}\left(J_2+J_3- 2\sqrt{(J_2-J_3)^2 + J_1^2}\right)$.
The interactions are set to $J_2 = -1$ and $J_3 = 1$, and $J_1 = \alpha$ is a variable. This leads 
to the lowest two eigenvalues: $E_0 (S^z=0) = -\frac{3}{2} \alpha$ and $E_0(S^z = 1) = -\frac{1}{2}\sqrt{4+\alpha^2}$.
As we consider the $\alpha > 0$ case only, the critical value $\alpha_c = \frac{1}{\sqrt{2}}$ for this toy model.


\section*{References}

\begin{thebibliography}{10}
\expandafter\ifx\csname url\endcsname\relax
  \def\url#1{\texttt{#1}}\fi
\expandafter\ifx\csname urlprefix\endcsname\relax\def\urlprefix{URL }\fi
\expandafter\ifx\csname href\endcsname\relax
  \def\href#1#2{#2} \def\path#1{#1}\fi

\bibitem{white1994}
S.~R. White, R.~M. Noack, D.~J. Scalapino, Resonating valence bond theory of
  coupled heisenberg chains, Phys. Rev. Lett. 73 (1994) 886--889.
\newblock \href {http://dx.doi.org/10.1103/PhysRevLett.73.886}
  {\path{doi:10.1103/PhysRevLett.73.886}}.

\bibitem{mk2015}
M.~Kumar, A.~Parvej, Z.~G. Soos, Level crossing, spin structure factor and
  quantum phases of the frustrated spin-1/2 chain with first and second
  neighbor exchange, J. Phys.: Condens. Matter 27~(31) (2015) 316001.
\newblock \href {http://dx.doi.org/10.1088/0953-8984/27/31/316001}
  {\path{doi:10.1088/0953-8984/27/31/316001}}.

\bibitem{verkholyak2012}
T.~Verkholyak, J.~Strečka, Quantum phase transitions in the exactly solved
  spin-1/2 heisenberg–ising ladder, Journal of Physics A: Mathematical and
  Theoretical 45~(30) (2012) 305001.
\newblock \href {http://dx.doi.org/10.1088/1751-8121/45/30/305001}
  {\path{doi:10.1088/1751-8121/45/30/305001}}.

\bibitem{Honecker2000}
A.~Honecker, F.~Mila, M.~Troyer, Magnetization plateaux and jumps in a class of
  frustrated ladders: A simple route to a complex behaviour, The European
  Physical Journal B - Condensed Matter and Complex Systems" 15~(2) (2000)
  227--233.
\newblock \href {http://dx.doi.org/10.1007/s100510051120}
  {\path{doi:10.1007/s100510051120}}.

\bibitem{sandvik1995}
A.~W. Sandvik, E.~Dagotto, D.~J. Scalapino, Spin dynamics of
  sr${\mathrm{cu}}_{2}$${\mathrm{o}}_{3}$ and the heisenberg ladder, Phys. Rev.
  B 53 (1996) R2934--R2937.
\newblock \href {http://dx.doi.org/10.1103/PhysRevB.53.R2934}
  {\path{doi:10.1103/PhysRevB.53.R2934}}.

\bibitem{dcjhon1987}
D.~C. Johnston, J.~W. Johnson, D.~P. Goshorn, A.~J. Jacobson, Magnetic
  susceptibility of (vo)$_2$p$_2$o$_7$: A one-dimensional spin-1/2 heisenberg
  antiferromagnet with a ladder spin configuration and a singlet ground state,
  Phys. Rev. B 35 (1987) 219--222.
\newblock \href {http://dx.doi.org/10.1103/PhysRevB.35.219}
  {\path{doi:10.1103/PhysRevB.35.219}}.

\bibitem{dagotto96}
E.~Dagotto, T.~M. Rice, Surprises on the way from one- to two-dimensional
  quantum magnets: The ladder materials, Science 271~(5249) (1996) 618--623.
\newblock \href {http://dx.doi.org/10.1126/science.271.5249.618}
  {\path{doi:10.1126/science.271.5249.618}}.

\bibitem{Maeshima2003}
N.~Maeshima, M.~Hagiwara, Y.~Narumi, K.~Kindo, T.~C. Kobayashi, K.~Okunishi,
  \href{http://stacks.iop.org/0953-8984/15/i=21/a=309}{Magnetic properties of a
  s = 1/2 zigzag spin chain compound (n 2 h 5 )cucl 3}, Journal of Physics:
  Condensed Matter 15~(21) (2003) 3607.
\newline\urlprefix\url{http://stacks.iop.org/0953-8984/15/i=21/a=309}

\bibitem{dutton_prl}
S.~E. Dutton, M.~Kumar, M.~Mourigal, Z.~G. Soos, J.-J. Wen, C.~L. Broholm,
  N.~H. Andersen, Q.~Huang, M.~Zbiri, R.~Toft-Petersen, R.~J. Cava, Quantum
  spin liquid in frustrated one-dimensional ${\mathrm{licusbo}}_{4}$, Phys.
  Rev. Lett. 108 (2012) 187206.
\newblock \href {http://dx.doi.org/10.1103/PhysRevLett.108.187206}
  {\path{doi:10.1103/PhysRevLett.108.187206}}.

\bibitem{Mourigal2012}
M.~Mourigal, M.~Enderle, B.~F\aa{}k, R.~K. Kremer, J.~M. Law, A.~Schneidewind,
  A.~Hiess, A.~Prokofiev, Evidence of a bond-nematic phase in
  ${\mathrm{licuvo}}_{4}$, Phys. Rev. Lett. 109 (2012) 027203.
\newblock \href {http://dx.doi.org/10.1103/PhysRevLett.109.027203}
  {\path{doi:10.1103/PhysRevLett.109.027203}}.

\bibitem{Drechsler2007}
S.-L. Drechsler, O.~Volkova, A.~N. Vasiliev, N.~Tristan, J.~Richter,
  M.~Schmitt, H.~Rosner, J.~M\'alek, R.~Klingeler, A.~A. Zvyagin, B.~B\"uchner,
  Frustrated cuprate route from antiferromagnetic to ferromagnetic
  spin-$\frac{1}{2}$ heisenberg chains: ${\mathrm{li}}_{2}{\mathrm{zrcuo}}_{4}$
  as a missing link near the quantum critical point, Phys. Rev. Lett. 98 (2007)
  077202.
\newblock \href {http://dx.doi.org/10.1103/PhysRevLett.98.077202}
  {\path{doi:10.1103/PhysRevLett.98.077202}}.

\bibitem{dagotto92}
E.~Dagotto, J.~Riera, D.~Scalapino, Superconductivity in ladders and coupled
  planes, Phys. Rev. B 45 (1992) 5744--5747.
\newblock \href {http://dx.doi.org/10.1103/PhysRevB.45.5744}
  {\path{doi:10.1103/PhysRevB.45.5744}}.

\bibitem{hijii2005}
K.~Hijii, A.~Kitazawa, K.~Nomura, Phase diagram of $\mathrm{S}=\frac{1}{2}$
  two-leg $xxz$ spin-ladder systems, Phys. Rev. B 72 (2005) 014449.
\newblock \href {http://dx.doi.org/10.1103/PhysRevB.72.014449}
  {\path{doi:10.1103/PhysRevB.72.014449}}.

\bibitem{vekua2003}
T.~Vekua, G.~I. Japaridze, H.-J. Mikeska, Phase diagrams of spin ladders with
  ferromagnetic legs, Phys. Rev. B 67 (2003) 064419.
\newblock \href {http://dx.doi.org/10.1103/PhysRevB.67.064419}
  {\path{doi:10.1103/PhysRevB.67.064419}}.

\bibitem{almeida2007}
J.~Almeida, M.~A. Martin-Delgado, G.~Sierra, Density-matrix renormalization
  group study of the bond-alternating $s=1∕2$ heisenberg ladder with
  ferro-antiferromagnetic couplings, Phys. Rev. B 76 (2007) 184428.
\newblock \href {http://dx.doi.org/10.1103/PhysRevB.76.184428}
  {\path{doi:10.1103/PhysRevB.76.184428}}.

\bibitem{dutton2012}
S.~E. Dutton, M.~Kumar, Z.~G. Soos, C.~L. Broholm, R.~J. Cava, Dominant
  ferromagnetism in the spin-1/2 half-twist ladder 334 compounds,
  ba${}_{3}$cu${}_{3}$in${}_{4}$o${}_{12}$ and
  ba${}_{3}$cu${}_{3}$sc${}_{4}$o${}_{12}$, J. Phys.: Condens. Matter 24~(16)
  (2012) 166001.
\newblock \href {http://dx.doi.org/10.1088/0953-8984/24/16/166001}
  {\path{doi:10.1088/0953-8984/24/16/166001}}.

\bibitem{mk_jpcm2013}
M.~Kumar, S.~E. Dutton, R.~J. Cava, Z.~G. Soos, Spin-flop and antiferromagnetic
  phases of the ferromagnetic half-twist ladder compounds
  ba${}_{3}$cu${}_{3}$in${}_{4}$o${}_{12}$ and
  ba${}_{3}$cu${}_{3}$sc${}_{4}$o${}_{12}$, J. Phys.: Condens. Matter 25~(13)
  (2013) 136004.
\newblock \href {http://dx.doi.org/10.1088/0953-8984/25/13/136004}
  {\path{doi:10.1088/0953-8984/25/13/136004}}.

\bibitem{volkova2012}
O.~S. Volkova, I.~S. Maslova, R.~Klingeler, M.~Abdel-Hafiez, Y.~C. Arango,
  A.~U.~B. Wolter, V.~Kataev, B.~B\"uchner, A.~N. Vasiliev, Orthogonal spin
  arrangement as possible ground state of three-dimensional shastry-sutherland
  network in ba${}_{3}$cu${}_{3}$in${}_{4}$o${}_{12}$, Phys. Rev. B 85 (2012)
  104420.
\newblock \href {http://dx.doi.org/10.1103/PhysRevB.85.104420}
  {\path{doi:10.1103/PhysRevB.85.104420}}.

\bibitem{volkova2016}
D.~I. Badrtdinov, O.~S. Volkova, A.~A. Tsirlin, I.~V. Solovyev, A.~N. Vasiliev,
  V.~V. Mazurenko, Hybridization and spin-orbit coupling effects in the
  quasi-one-dimensional spin-$\frac{1}{2}$ magnet
  ${\mathrm{ba}}_{3}{\mathrm{cu}}_{3}{\mathrm{sc}}_{4}{\mathrm{o}}_{12}$, Phys.
  Rev. B 94 (2016) 054435.
\newblock \href {http://dx.doi.org/10.1103/PhysRevB.94.054435}
  {\path{doi:10.1103/PhysRevB.94.054435}}.

\bibitem{mk2016}
Z.~G. Soos, A.~Parvej, M.~Kumar, Numerical study of incommensurate and
  decoupled phases of spin-1/2 chains with isotropic exchange $j_1$ , $j_2$
  between first and second neighbors, J. Phys.: Condens. Matter 28~(17) (2016)
  175603.
\newblock \href {http://dx.doi.org/10.1088/0953-8984/28/17/175603}
  {\path{doi:10.1088/0953-8984/28/17/175603}}.

\bibitem{mk2013}
M.~Kumar, Z.~G. Soos, Decoupled phase of frustrated spin-$\frac{1}{2}$
  antiferromagnetic chains with and without long-range order in the ground
  state, Phys. Rev. B 88 (2013) 134412.
\newblock \href {http://dx.doi.org/10.1103/PhysRevB.88.134412}
  {\path{doi:10.1103/PhysRevB.88.134412}}.

\bibitem{chitra95}
R.~Chitra, S.~Pati, H.~R. Krishnamurthy, D.~Sen, S.~Ramasesha, Density-matrix
  renormalization-group studies of the spin-1/2 heisenberg system with
  dimerization and frustration, Phys. Rev. B 52 (1995) 6581--6587.
\newblock \href {http://dx.doi.org/10.1103/PhysRevB.52.6581}
  {\path{doi:10.1103/PhysRevB.52.6581}}.

\bibitem{ckm69b}
C.~K. Majumdar, D.~K. Ghosh, On next‐nearest‐neighbor interaction in linear
  chain. ii, J. Math. Phys. 10~(8) (1969) 1399--1402.
\newblock \href {http://dx.doi.org/10.1063/1.1664979}
  {\path{doi:10.1063/1.1664979}}.

\bibitem{suhl98}
H.~Suhl, I.~K. Schuller, Spin-wave theory of exchange-induced anisotropy, Phys.
  Rev. B 58 (1998) 258--264.
\newblock \href {http://dx.doi.org/10.1103/PhysRevB.58.258}
  {\path{doi:10.1103/PhysRevB.58.258}}.

\bibitem{hong98}
T.~M. Hong, Simple mechanism for a positive exchange bias, Phys. Rev. B 58
  (1998) 97--100.
\newblock \href {http://dx.doi.org/10.1103/PhysRevB.58.97}
  {\path{doi:10.1103/PhysRevB.58.97}}.

\bibitem{cpbean56}
W.~H. Meiklejohn, C.~P. Bean, New magnetic anisotropy, Phys. Rev. 102 (1956)
  1413--1414.
\newblock \href {http://dx.doi.org/10.1103/PhysRev.102.1413}
  {\path{doi:10.1103/PhysRev.102.1413}}.

\bibitem{cpbean57}
W.~H. Meiklejohn, C.~P. Bean, New magnetic anisotropy, Phys. Rev. 105 (1957)
  904--913.
\newblock \href {http://dx.doi.org/10.1103/PhysRev.105.904}
  {\path{doi:10.1103/PhysRev.105.904}}.

\bibitem{nogues99}
J.~Nogués, I.~K. Schuller, Exchange bias, J. Magn. Magn. Mater. 192~(2) (1999)
  203 -- 232.
\newblock \href {http://dx.doi.org/10.1016/S0304-8853(98)00266-2}
  {\path{doi:10.1016/S0304-8853(98)00266-2}}.

\bibitem{kiwi2001}
M.~Kiwi, Exchange bias theory, J. Magn. Magn. Mater. 234~(3) (2001) 584 -- 595.
\newblock \href {http://dx.doi.org/10.1016/S0304-8853(01)00421-8}
  {\path{doi:10.1016/S0304-8853(01)00421-8}}.

\bibitem{berkowitz99}
A.~Berkowitz, K.~Takano, Exchange anisotropy -- a review, J. Magn. Magn. Mater.
  200~(1–3) (1999) 552 -- 570.
\newblock \href {http://dx.doi.org/10.1016/S0304-8853(99)00453-9}
  {\path{doi:10.1016/S0304-8853(99)00453-9}}.

\bibitem{finazzi2004}
M.~Finazzi, Interface coupling in a ferromagnet/antiferromagnet bilayer, Phys.
  Rev. B 69 (2004) 064405.
\newblock \href {http://dx.doi.org/10.1103/PhysRevB.69.064405}
  {\path{doi:10.1103/PhysRevB.69.064405}}.

\bibitem{malozemoff87}
A.~P. Malozemoff, Random-field model of exchange anisotropy at rough
  ferromagnetic-antiferromagnetic interfaces, Phys. Rev. B 35 (1987)
  3679--3682.
\newblock \href {http://dx.doi.org/10.1103/PhysRevB.35.3679}
  {\path{doi:10.1103/PhysRevB.35.3679}}.

\bibitem{koon97}
N.~C. Koon, Calculations of exchange bias in thin films with
  ferromagnetic/antiferromagnetic interfaces, Phys. Rev. Lett. 78 (1997)
  4865--4868.
\newblock \href {http://dx.doi.org/10.1103/PhysRevLett.78.4865}
  {\path{doi:10.1103/PhysRevLett.78.4865}}.

\bibitem{schulthess98}
T.~C. Schulthess, W.~H. Butler, Consequences of spin-flop coupling in exchange
  biased films, Phys. Rev. Lett. 81 (1998) 4516--4519.
\newblock \href {http://dx.doi.org/10.1103/PhysRevLett.81.4516}
  {\path{doi:10.1103/PhysRevLett.81.4516}}.

\bibitem{schulthess99}
T.~C. Schulthess, W.~H. Butler, Coupling mechanisms in exchange biased films
  (invited), J. Appl. Phys. 85~(8) (1999) 5510--5515.
\newblock \href {http://dx.doi.org/10.1063/1.369878}
  {\path{doi:10.1063/1.369878}}.

\bibitem{kim2001}
J.-V. Kim, R.~L. Stamps, Defect-modified exchange bias, Applied Physics Letters
  79~(17) (2001) 2785--2787.
\newblock \href {http://dx.doi.org/10.1063/1.1413731}
  {\path{doi:10.1063/1.1413731}}.

\bibitem{nowak2001}
U.~Nowak, A.~Misra, K.~D. Usadel, Domain state model for exchange bias, J.
  Appl. Phys. 89~(11) (2001) 7269--7271.
\newblock \href {http://dx.doi.org/10.1063/1.1358829}
  {\path{doi:10.1063/1.1358829}}.

\bibitem{sakurai93}
Y.~Sakurai, H.~Kitatani, T.~Ishiguro, Y.~Ichinose, N.~S. Kazama, Monte carlo
  approach for f-af coupling layers, IEEE Trans. Magn. 29 (1993) 3879.

\bibitem{mauri87}
D.~Mauri, H.~C. Siegmann, P.~S. Bagus, E.~Kay, Simple model for thin
  ferromagnetic films exchange coupled to an antiferromagnetic substrate, J.
  Appl. Phys. 62~(7) (1987) 3047--3049.
\newblock \href {http://dx.doi.org/10.1063/1.339367}
  {\path{doi:10.1063/1.339367}}.

\bibitem{xi99}
H.~Xi, R.~M. White, S.~M. Rezende, Irreversible and reversible measurements of
  exchange anisotropy, Phys. Rev. B 60 (1999) 14837--14840.
\newblock \href {http://dx.doi.org/10.1103/PhysRevB.60.14837}
  {\path{doi:10.1103/PhysRevB.60.14837}}.

\bibitem{xi2000}
H.~Xi, R.~M. White, A theoretical study of interfacial spin flop in
  exchange-coupled bilayers, IEEE Trans. Magn. 36 (2000) 2635.

\bibitem{geshev2000}
J.~Geshev, Analytical solutions for exchange bias and coercivity in
  ferromagnetic/antiferromagnetic bilayers, Phys. Rev. B 62 (2000) 5627--5633.
\newblock \href {http://dx.doi.org/10.1103/PhysRevB.62.5627}
  {\path{doi:10.1103/PhysRevB.62.5627}}.

\bibitem{schulthess98b}
T.~C. Schulthess, W.~H. Butler, First-principles exchange interactions between
  ferromagnetic and antiferromagnetic films: Co on nimn, a case study, J. Appl.
  Phys. 83~(11) (1998) 7225--7227.
\newblock \href {http://dx.doi.org/10.1063/1.367824}
  {\path{doi:10.1063/1.367824}}.

\bibitem{white-prl92}
S.~R. White, Density matrix formulation for quantum renormalization groups,
  Phys. Rev. Lett. 69 (1992) 2863--2866.
\newblock \href {http://dx.doi.org/10.1103/PhysRevLett.69.2863}
  {\path{doi:10.1103/PhysRevLett.69.2863}}.

\bibitem{karen2006}
K.~A. Hallberg, New trends in density matrix renormalization, Advances in
  Physics 55~(5-6) (2006) 477--526.

\bibitem{schollwock2005}
U.~Schollw\"ock, The density-matrix renormalization group, Rev. Mod. Phys. 77
  (2005) 259--315.

\bibitem{mk2010b}
M.~Kumar, Z.~G. Soos, D.~Sen, S.~Ramasesha, Modified density matrix
  renormalization group algorithm for the zigzag spin-$\frac{1}{2}$ chain with
  frustrated antiferromagnetic exchange: Comparison with field theory at large
  ${J}_{2}/{J}_{1}$, Phys. Rev. B 81 (2010) 104406.
\newblock \href {http://dx.doi.org/10.1103/PhysRevB.81.104406}
  {\path{doi:10.1103/PhysRevB.81.104406}}.

\bibitem{ddpbc2016}
D.~Dey, D.~Maiti, M.~Kumar, An efficient density matrix renormalization group
  algorithm for chains with periodic boundary condition, Papers in Physics 8
  (2016) 080006.

\bibitem{fulde95}
P.~Fulde, Electron Correlations in Molecules and Solids, 3rd Edition,
  Springer-Verlag, 1995.

\bibitem{mila98}
F.~Mila, Ladders in a magnetic field: a strong coupling approach, Eur. Phys. J.
  B 6~(2) (1998) 201--205.
\newblock \href {http://dx.doi.org/10.1007/s100510050542}
  {\path{doi:10.1007/s100510050542}}.

\bibitem{white96}
S.~R. White, I.~Affleck, Dimerization and incommensurate spiral spin
  correlations in the zigzag spin chain: Analogies to the kondo lattice, Phys.
  Rev. B 54 (1996) 9862--9869.
\newblock \href {http://dx.doi.org/10.1103/PhysRevB.54.9862}
  {\path{doi:10.1103/PhysRevB.54.9862}}.

\end{thebibliography}
\end{document}